\documentclass[10pt]{iopart}

\usepackage{graphicx}
\usepackage{amsmath}
\begin{document}

\title[Collision frequency measurement with the hairpin resonator probe]{Electron neutral collision frequency measurement with the hairpin resonator probe}

\author{David J Peterson$^{[1]}$, Philip Kraus$^{[2]}$, Thai Cheng Chua$^{[2]}$, Lynda Larson$^{[3]}$, Steven C Shannon$^{[1]}$}

\address{[1] North Carolina State University, Nuclear Engineering Department, Raleigh NC, USA}
\address{[2] Applied Materials, Sunnyvale CA, USA}
\address{[3] Treasure Isle Jewelers, Cary NC, USA}
\ead{djpeter5@ncsu.edu}
\vspace{10pt}
\begin{indented}
\item[]March 2017
\end{indented}

\begin{abstract}
Electron neutral collision frequency is measured using both grounded and floating hairpin resonator probes in a 27 MHz parallel plate capacitively coupled plasma (CCP). Operating conditions are 0.1-2 Torr (13.3-267 Pa) in Ar, He, and Ar-He gas mixtures. The method treats the hairpin probe as a two wire transmission line immersed in a dielectric medium. A minimization method is applied during the pressure and sheath correction process by sweeping over assumed collision frequencies in order to obtain the measured collision frequency. Results are compared to hybrid plasma equipment module (HPEM) simulations and show good agreement. 
\end{abstract}

%
%
%
%
\ioptwocol

\section{Introduction}
Moderate pressure plasmas ($1-10$ Torr) are an area of growing interest as they have potential for faster processing, and are especially useful for thin film deposition[1]. Standard diagnostic approaches like Langmuir probes have formidable complications at these pressures due to the complex nature of collisional sheath dynamics[2]. Given the lack of easily implemented localized measurements at these pressures, any extra information that can be obtained is useful for improving industrial source design and validation efforts for plasma chemistry models.

The hairpin resonance probe operates on a similar principle to the cavity perturbation technique. Both rely on the resonance frequency shift induced by the plasma's lossy dielectric properties to infer electron density ($n_{e}$), while the hairpin probe has the added benefit of allowing localized measurements. It has been previously suggested that resonance broadening can be used to directly determine the electron-neutral collision frequency[3-4]. This work extends the hairpin probe's capabilities, allowing measurement of electron neutral collision frequency ($\nu_{en}$) using a relationship between hairpin probe resonance broadening and electron neutral collisions. 

\section{Hairpin Theory}
The vacuum resonant frequency of a hairpin resonator, with tine lengths $l$, is given by $f_{0} = c/4l$, where $c$ is the vacuum speed of light. In a plasma, the resonance shifts to a higher frequency $f_{r} = c/4l\sqrt{\epsilon'}$, where $\epsilon'$ is the real part of the complex plasma permittivity 
\begin{equation} \label{eq: plasma permittivity}
\epsilon_{p} = 1 - \frac{\omega_{p}^{2}}{\omega_{r}\left( \omega_{r} - i\nu_{en}\right) }.
\end{equation}

The previous relations, the angular resonant frequency $\omega_{r} = 2\pi f_{r}$, and the electron plasma frequency $\omega_{p} = (n_{e}e^{2}/m_{e}\epsilon_{0})^{1/2}$ can be combined with measurements and a simple result from transmission line theory to produce a closed system of equations where $\nu_{en}$ is the only unknown. 

This approach implicitly assumes $n_{e}$ is accurately determined by the hairpin probe. A value for electron temperature ($T_{e}$) must be assumed to correct for the presence of a sheath around the hairpin tines. Since the probe is not biased, sheaths act primarily as a geometric effect pertaining to the volume of free space between the hairpin width ($w$). Experimental error introduced by uncertainty in the sheath width ($b$) is therefore a quantity subject to optimization through probe design, further discussed in section 3.2.

Sheath and pressure corrections must be applied to accurately determine $n_{e}$. The sheath correction factor is the same used by Sands et al [5], shown in equation (\ref{eq: sheath correction}). 
\begin{equation} \label{eq: sheath correction}
\xi_{s} = 1 - \frac{f_{0}^{2}}{f_{r}^{2}}\frac{\left[ \ln\left(\frac{w-a}{w-b} \right) + \ln\left(\frac{b}{a} \right)  \right] }{\ln\left( \frac{w-a}{a}\right) }
\end{equation} 
The correction is applied using the iterative approach developed by Piejak [6]. The sheath is assumed to extend one electron Debye length ($\lambda_{D} = \epsilon_{0}k_{B}T_{e}/e^{2}n_{e}$) out from the radius of the hairpin tine ($a$) for both types of probes. Here $\epsilon_{0}$, $k_{B}$, and $e$ are the permittivity of free space, boltzmann's constant, and electronic charge, respectively. Measurements in this work are made with both floating and grounded hairpin probes. This is done primarily to quantify the error introduced by using grounded probes, as opposed to measurements made with a floating probe which only require a DC sheath correction [7]. Differences between floating and grounded probes are discussed in section 3.2.

A pressure correction factor ($\xi_{p}$) must be applied at moderate pressures in order to correct for the centerline shift in the resonance frequency. This is done using the expression given by Sands,

\begin{equation} \label{eq: pressure correction}
\xi_{p} = \frac{1}{1 + \left(\frac{\nu_{en}}{2\pi f_{0}} \right)^{2}} .
\end{equation}
The pressure and sheath corrections update $n_{e}$ in the manner shown in equation (\ref{eq: electron density}). The pressure correction is applied first and then sent to the sheath correction, which iteratively solves for the corrected $n_{e}$.

\begin{equation} \label{eq: electron density}
n_{e} = \frac{\pi m_{e}}{e^{2}}\frac{f_{r2}^{2} - f_{0}^{2}}{\xi_{p}\xi_{s}}
\end{equation}

The simplest approach to determining $\nu_{en}$ is to treat the plasma as a lossy dielectric. The plasma quality factor ($Q_{Plasma}$) can be defined using the ratio of the real and complex plasma permittivity [8], as seen in equation ($\ref{eq: q plasma}$). 
\begin{equation} \label{eq: q plasma}
Q_{Plasma} = \frac{\epsilon'}{\epsilon''} = \frac{1-\frac{\omega_{p}^{2}}{\omega_{r}^{2} + \nu_{en}^{2}}}{\frac{\nu_{en}}{\omega_{r}}\frac{\omega_{p}^{2}}{\omega_{r}^{2}+\nu_{en}^{2}}}
\end{equation}
The same result can also be obtained from transmission line theory, where the attenuation constant ($\alpha$) is defined using distributed parameters and assumes negligible resistive losses in the probe. This is a valid assumption considering the conducting material of the probe is silver. The analysis assumes weak attenuation, where $\alpha l \ll 1$, also a valid assumption for the pressure regimes being investigated. A complete description of transmission line analysis of the hairpin probe can be found by Xu et al [4].

When the probe is immersed in plasma, the hairpin can be treated as a loaded resonant quarter wave transmission line[9], resulting in the coupling of Q values in the manner shown by equation (\ref{eq: q measured}). 
\begin{equation} \label{eq: q measured}
\frac{1}{Q_{Measured}} = \frac{1}{Q_{Vacuum}} + \frac{1}{Q_{Plasma}}
\end{equation}
$Q_{vacuum}$ is the measured quality factor of the hairpin inside the chamber at vacuum. Measurements are made inside the chamber in order to completely account for parasitic loading of chamber components. 

Solving for $Q_{Plasma}$ using equation (\ref{eq: q measured}), we see that it now consists entirely of measured quantities. Both $Q_{Vacuum}$ and $Q_{Measured}$ are measured using $Q=f_{r}/\Delta f_{r}$, where $\Delta f_{r}$ is the full width half max (FWHM) of the Lorentzian resonance profile. Collision frequency can then be determined by combining equations (\ref{eq: q plasma}) and (\ref{eq: q measured}), since $\omega_{r}$ and $\omega_{p}$ are both known.

\section{Experimental}
\subsection{Experimental Setup}

Experiments are performed on the Modular Radiofrequency Plasma Chamber (MrPC). MrPC is a parallel plate CCP reactor with two 150 mm diameter aluminum electrodes. The electrodes are mechanically fastened to spindles for adjusting the distance between the electrodes and their position inside the reactor. The electrode gap was 1.91 cm for all measurements. The electrodes are housed in a Rexolite plastic shroud which provides 180 pF of electrical isolation to the surrounding ground plate. One electrode is powered and the other grounded. The powered electrode is connected to a matching network through RG-393 MIL-C-17 RF cable and terminated on each end with N-Type connectors. MrPC is a stainless steel vessel pumped by a turbomolecular pump (TMP) with a base pressure $<$0.2 mTorr. The leak rate of the reactor when isolated from the TMP was 7.6 mTorr/min. The gas introduced by the leak is negligible compared to the controlled gas flow for all experiments. Gas is fed to the chamber using an analog mass flow controller (MFC) and pressure is controlled with a closed-loop capacitance manometer and throttle valve. Operating conditions are 0.1-2 Torr in Ar, He, and Ar-He gas mixtures powered by a 27.12 MHz LVG RF generator from ENI Products. A computer controlled HP8753C network analyzer is used for probe data acquisition. A diagram of the experimental setup is shown in figure 1. A more complete description of the setup has been given previously by Zhang et al[10].

\begin{figure}[h]
\begin{center}
\includegraphics[scale=0.5]{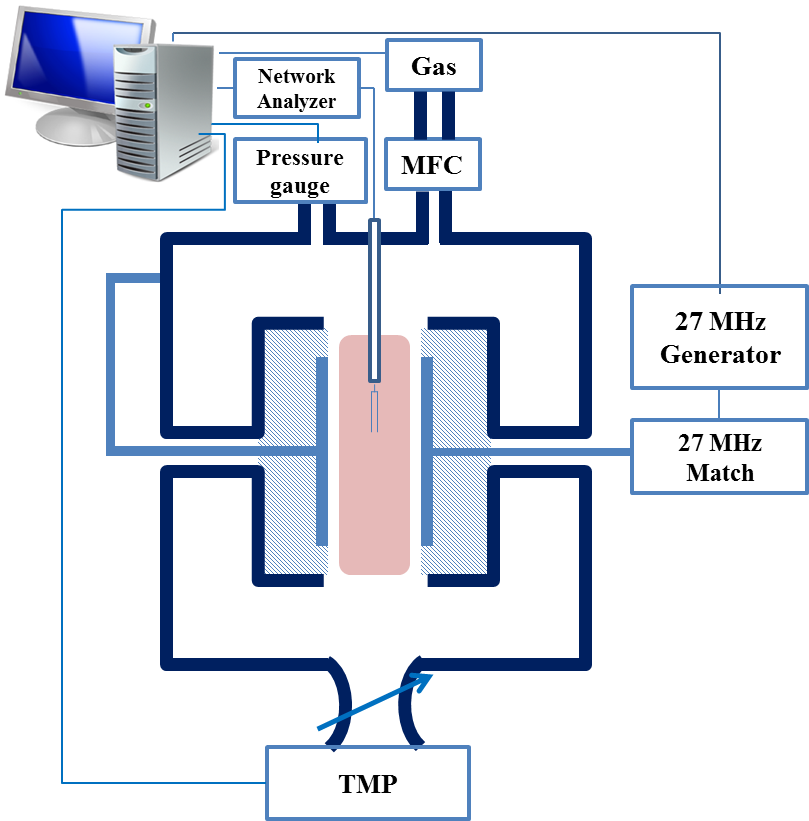}
\end{center}
\caption{Experimental setup for measurements in the CCP parallel plate configuration.}
\end{figure}

\subsection{Hairpin Design and Calibration}

Grounded hairpin probes provide a simple design for producing high Q resonances, which are necessary for measurements at moderate pressures. The high Q resonances are a result of direct electrical connection between the driving loop and the hairpin tines. Grounded probes come with a caveat. They require more sophisticated sheath analysis due to the presence of a grounded RF sheath across wire surfaces. 

Floating hairpin probes are used in this experiment to simplify the sheath correction process, only requiring DC sheath corrections. The floating probe design used in this work is optimized to keep Q high while still isolating the hairpin from ground to ensure no RF driven sheath is formed. A diagram of the design is shown in figure 2. A quartz sleeve ensures a small capacitive impedance for the microwave frequencies sent to the probe, which are typically in the low GHz range. Meanwhile, the drive frequency (27 MHz) impedance remains large enough to inhibit the formation of a RF driven sheath. This is done by keeping the thickness of the quartz sleeve ($t$) relatively thin, around 1.0 mm. The floating probe used in this work had $t$=1.15 mm, which corresponds to a capacitive impedance approximately 15 $\Omega$ at 2 GHz. Based on observed $Q$ of this probe, a thinner quartz sleeve would further improve performance without introducing a significant RF sheath. Length of the metal ring ($L$) serves as an additional parameter for fine tuning of impedance by changing the effective area of the capacitor. Microwave plasmas can avoid this optimization step since RF driven sheaths will not exist with drive frequencies above the plasma frequency.

The floating and grounded probe wire radii ($a$) and width ($w$) used in this experiment were $a$=0.22 and 0.325 mm, and $w$=1.85 and 3.05 mm, respectively. All probe components were made of sterling silver, which was annealed and then drawn through a draw plate to the specified radii. Rounding pliers were used to shape the loop to the correct orientation. The loop and tines were then TIG welded to their respective positions. A drop of silver was deposited to the base of the hairpin tines before welding in order to avoid unintentionally destroying the tines. Metal notches were used to secure the position of the quartz sleeve for the floating probe.

\begin{figure}[h]
\begin{center}
\includegraphics[scale=0.3]{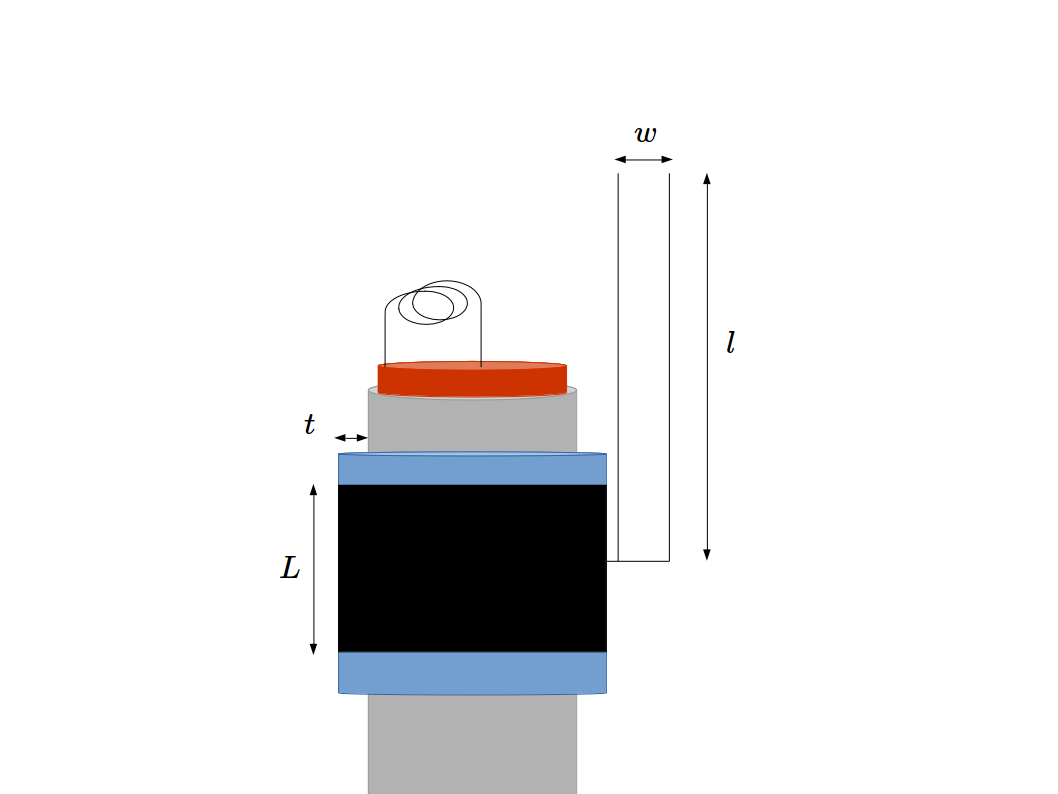}
\end{center}
\caption{Diagram of floating probe design. The inner grounded jacket is copper and the outer grounded jacket is Inconel. The normal direction to the center of the driving loop is coincident with the center of the hairpin tines, not shown for illustrative purposes. The hairpin tines, driving loop, and capacitive metal ring (shown in black) are all made of silver. The quartz sleeve is blue.}
\end{figure}

Hairpin resonances are fit with Lorentz curves since they can be characterized as a RLC resonant circuit [11]. Acquired data is fed into open source Python and Fityk[12] scripts for automated peak fitting and data analysis. These were developed in-house and are freely available upon request. The collision frequency measurement technique relies on accurate determination of the FWHM of the hairpin resonance in both vacuum and plasma. A calibration is required to ensure accuracy. The calibration removes transmission line effects of the probe, which can produce substantially different Q factors. An illustration of these differences is shown in figure 3. Figure 3a shows an uncalibrated curve which has the characteristic transmission line curve along with the resonance. Figure 3b shows the same curve with an applied calibration, obtained by subtracting the calibration curve from the original. Applying the calibration produces up to a $20\%$ difference in measured Q factors due to distortion of the resonance peak. 

Calibration curves are obtained by shorting the far end of the hairpin tines. In this case, copper tape was rolled up in way that produces two holes at each end, with one side being crimped, and the un-crimped side put around the open circuit end of the hairpin. This ensures that good electrical contact is made on both tines, shorting the open circuit, and fits well enough to not fall off during experiments. 

\begin{figure}[h]
\begin{center}
\includegraphics[scale=0.3]{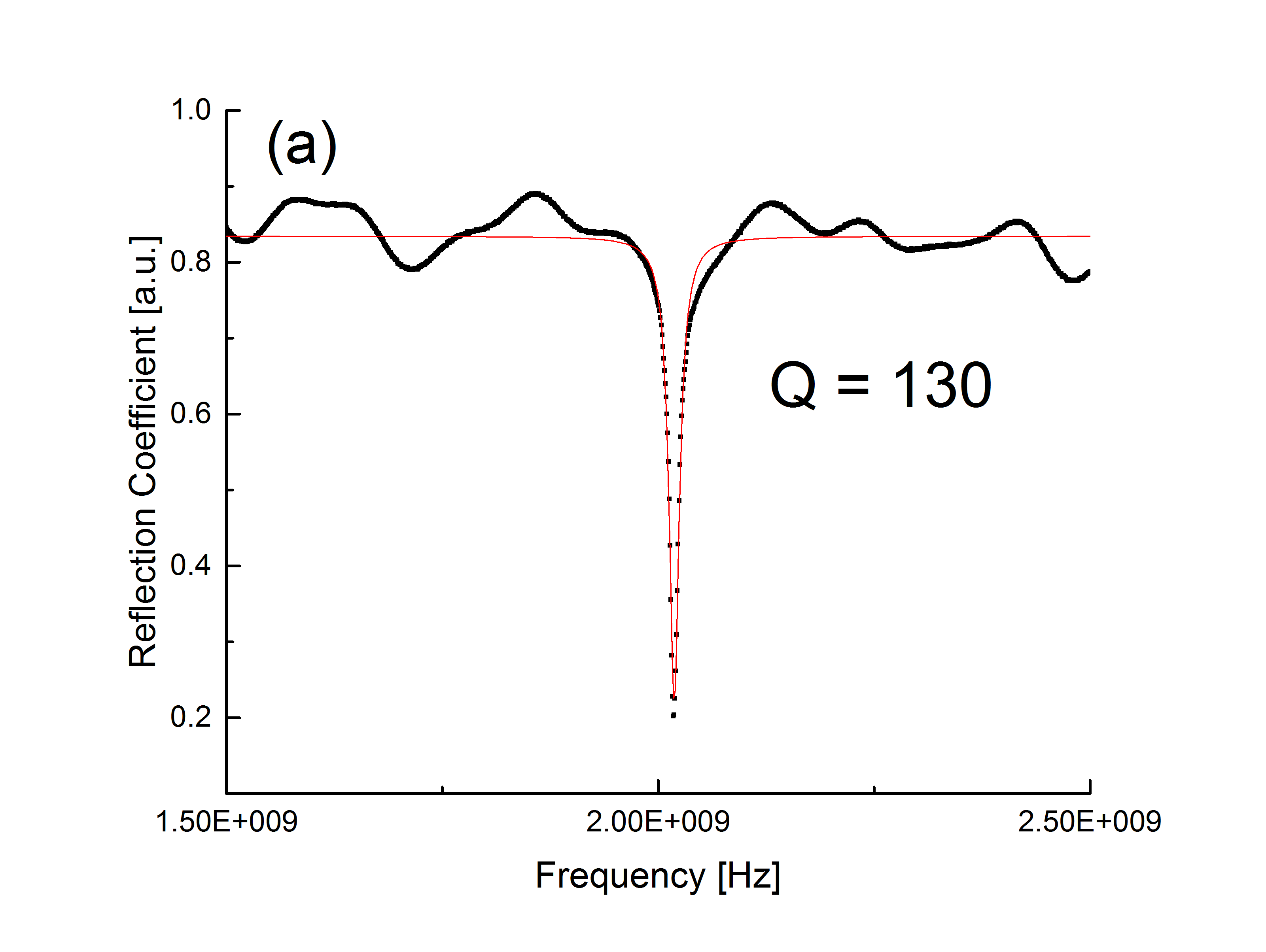}
\includegraphics[scale=0.3]{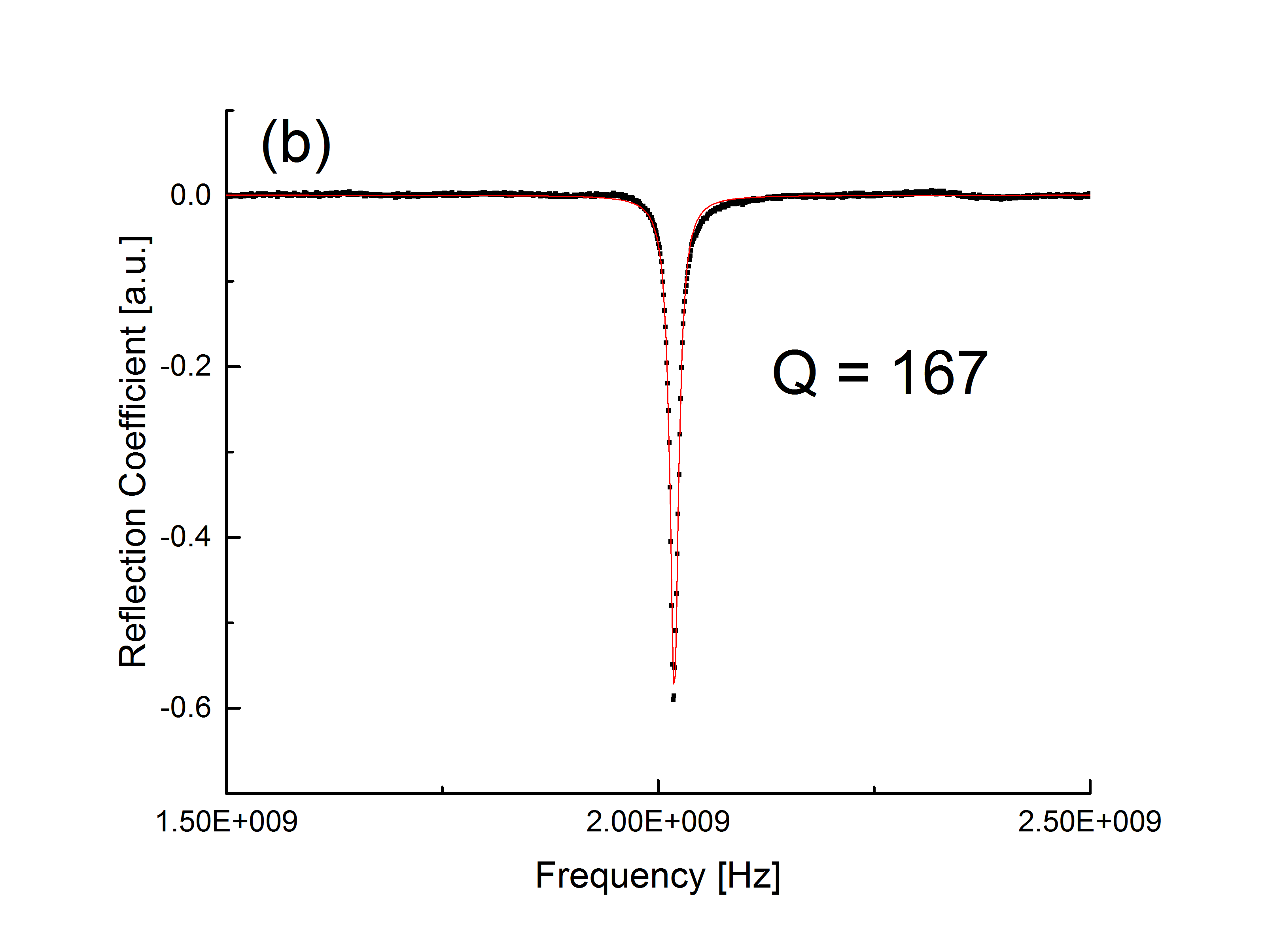}
\end{center}
\caption{(a) Uncalibrated reflection coefficient of hairpin resonance probe and a Lorentz fit showing clear transmission line properties. (b) The calibrated version of the same resonance with a fit. The uncalibrated curve in (a) is artificially broadened at the base due to the transmission line effect, resulting in a $20\%$ difference in measured quality factor.}
\end{figure}

Since transmission line properties shift when immersed in plasma, calibrations were taken for each unique set of plasma conditions in order to ensure accurate FWHM measurements. Calibrations obtained for different plasma conditions result in differences in Q that are typically less than the uncertainty resulting from the fit itself. Uncertainty in peak center and FWHM associated with fitting are typically both around 0.5\%. As a result, only one calibration is used for all plasma conditions. $Q_{Vacuum}$ is measured immediately after the plasma is extinguished. This ensures that ion heating of the probe is taken into account since it slightly decreases the quality factor and resonant frequency, a phenomenon previously noted by Piejak.

At higher pressures, one must also take steps to avoid probe perturbation of the plasma due to the drawing of excessive current. This begins to occur as the electron mean free path approaches the dimensions of the probe. The hairpin tine diameter is approximately 5 times larger than the electron mean free path at 1 Torr in argon, meaning that some plasma perturbation occurred during measurements. The perturbation presents a probe design optimization dilemma. Experimental design necessitates finding a balance between the desire for minimal perturbation and the need for a high enough Q factor to perform measurements at higher pressures.

Measurements are limited to conditions where electron density can accurately be measured. The smallest measurable $n_{e}$, suggested by Karkari et al [13], corresponds to a plasma frequency $f_{p} \sim (2/Q)^{1/2}f_{0}$. The probes used in this experiment have $Q_{Vacuum}\approx 380$, which corresponds to a lower density limit near $3 \times 10^{8}$ cm$^{-3}$. Measurements are also bound by an upper $n_{e}$ limit that stems from the degradation in signal quality. It is suggested that the vacuum resonant frequency be larger than the plasma frequency in order to avoid exciting waves in the plasma, which further perturbs local parameters and introduces an additional nonlinear loss mechanism. Avoiding this nonlinear loss mechanism is particularly important for this method. 

Pressure also acts as a limiting parameter to these measurements. At low pressures, in the 10 mTorr range, collisional broadening contributes a smaller fractional amount to the overall broadening and becomes harder to measure. The lowest measurable pressure for this experiment was 0.1 Torr. At higher pressures, in the 10 Torr range, resonances broaden significantly due to collisional damping. Novel probe designs with higher vacuum quality factors are capable of extending the measurable parameter space, and is the subject of future work.

\subsection{Determination of $\nu_{en}$ from Q}

In figure 4, $Q_{Measured}$ is fit with equation (\ref{eq: q measured}) by assuming a constant pressure normalized collision frequency $\nu_{en}/p$=2.5 GHz/Torr. The fit clearly captures the shape of $Q_{Measured}$, confirming expectations that pressure and sheath corrected $n_{e}$ plays an important role in determining resonance width. The peak in quality factor is accompanied by a drop in measured electron density, making the probe less lossy, which results in the unusual shape. Electron temperature can appreciably change over such a relatively large pressure range, contributing to deviations in the fit caused by assuming a constant pressure normalized collision frequency.

\begin{figure}[h]
\begin{center}
\includegraphics[scale=0.35]{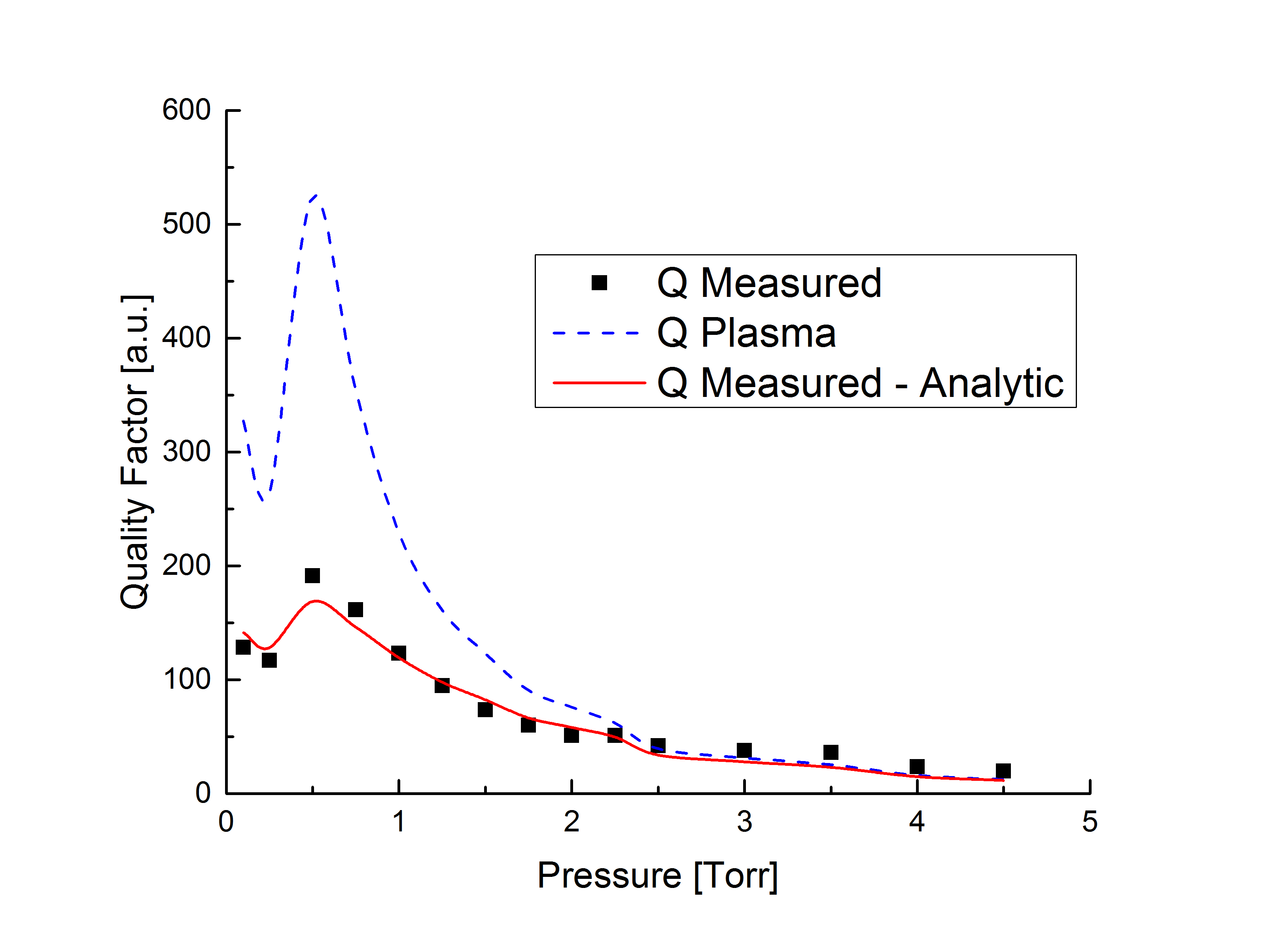}
\end{center}
\caption{Black squares show measured Q values over a range of pressures at constant applied power, $P$=100 W. Solid lines, with markers at pressures where measured values exist, are analytical fits. The red line with triangles is the fit given by using equation (\ref{eq: q plasma}) in equation (\ref{eq: q measured}), while the blue line with circles is only from equation (\ref{eq: q plasma}).}
\end{figure}
		
Instead of assuming the value of $\nu_{en}$, it can be solved for directly using equation (\ref{eq: q plasma}). However, the $n_{e}$ used for electron plasma frequency is a value that has already been modified by a $\nu_{en}$ dependent pressure correction. This apparent problem can be decoupled by sweeping over the initial collision frequency ($\nu_{en}^{i}$), producing a range of possible $n_{e}$. An example of this is illustrated in figure 5. The dark region corresponds to 1000 different pressure and sheath corrected density profiles, each assuming different normalized collision frequencies spanning 2-4 GHz. The unusual $n_{e}$ profile is a result of shifts in the spatial distribution of $n_{e}$ as pressure increases, a phenomenon also observed in simulations done for this work.   
		
\begin{figure}[h]
\begin{center}
\includegraphics[scale=0.55]{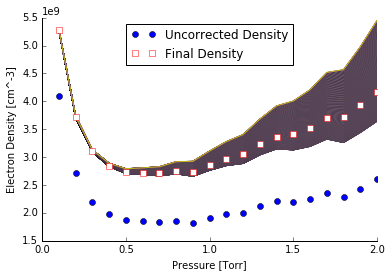}
\end{center}
\caption{Blue circles show uncorrected electron density for pure Ar and $P$=60W using a grounded hairpin probe. The darker region is 1000 different pressure and sheath corrected densities, each assuming different normalized collision frequencies spanning 2-4 GHz. The final corrected electron densities are given by red outlined white squares.}
\end{figure}  
				
Different $\nu_{en}^{i}$ that are used for the pressure correction yield different measured collision frequencies ($\nu_{en}^{m}$) for the same experimental conditions. An accurate pressure correction will minimize the difference between $\nu_{en}^{i}$ and $\nu_{en}^{m}$, which provides a path for the direct measurement of $\nu_{en}$.

\section{Results and Discussion}

\begin{figure}[h]
\begin{center}
\includegraphics[scale=0.55]{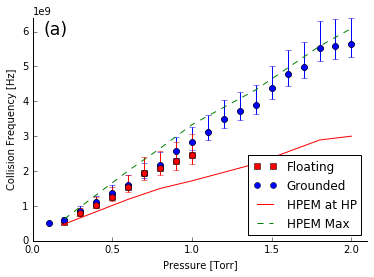}
\includegraphics[scale=0.55]{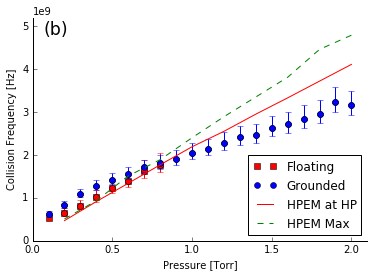}
\end{center}
\caption{(a) Measured collision frequency using the grounded hairpin probe in a pure argon discharge with $P$=60 W spanning $p$=0.1-2.0 Torr. Also included are the maximum and probe localized $\nu_{en}$ HPEM results over the same pressures. The sheath correction assumes $T_{e}$=3.2 eV. Error bars correspond to a sheath correction using 1 eV$<T_{e}<$5 eV and do not consider different sheath models. (b) Repeated measurements for pure helium at $P$=20 W. The sheath correction assumes $T_{e}$=4 eV. Error bars correspond to 2 eV $<T_{e}<$6 eV for the sheath correction.}
\end{figure}  

Floating and grounded probe measurements are taken in pure Ar over a pressure range of 0.1-2 Torr and compared with hybrid plasma equipment module [14] (HPEM) simulations in figure 6(a). The floating probe measurements are limited to a smaller pressure range of 0.2-1 Torr due to a lower $Q_{Vacuum}$. Measurements fall between the maximum (Max) and hairpin location specific (at HP) simulated $\nu_{en}$ produced by HPEM. The large range of simulated $\nu_{en}$ stems from the fact that Ar is a Ramsauer gas. This means that relatively small differences in $T_{e}$ can produce significantly different $\nu_{en}$. For example, the simulated $T_{e}$ that correspond to the maximum and at hairpin $\nu_{en}$ are approximately 1.8-3.3 eV, respectively. Floating probe measurements yielded smaller $\nu_{en}$ than the grounded probe measurements because floating probes only exhibit DC sheaths. The sheath on a floating probe will always be smaller than that of the grounded probe due to the rectification of RF current that occurs with grounded probes. The smaller sheath of the floating probe will result in a larger, and thus more accurate, uncorrected $n_{e}$. 

If identical sheath corrections are applied to both cases, as done in figures 6 (a) and (b), one would expect the larger $n_{e}$ from the floating probe to yield smaller corresponding $\nu_{en}$, easily seen in equation (\ref{eq: q plasma}). Sheath corrections assume $T_{e}$=3.2 eV, a typical value simulated in HPEM. Error bars correspond to sheath corrections with 1 eV $<T_{e}<$5 eV, shown primarily to capture rough uncertainty values resulting from an unknown $T_{e}$. HPEM simulations for pure Ar used an applied voltage of 55 V, corresponding to an simulated input power of 60 W, when assuming a matched 50 $\Omega$ load. Applied voltage was kept constant for all simulations, as opposed to letting voltage vary to reach a specified input power. Fixing applied voltage produced more consistent results and is recommended practice.

Measurements are made in pure He with $P$=20 W in figure 6(b). Floating and grounded probe $\nu_{en}$ values start to diverge near 0.2 Torr in the expected directions for the same reason previously mentioned. Floating probe measurements were again limited to 0.1-0.8 Torr because $Q_{vacuum}$ was too low. The measurable pressure range for the floating probe in He was slightly lower than for Ar because He measurements yielded higher $n_{e}$ and $\nu_{en}$ for the same conditions. This limited the measurable range due to signal loss. Sheath corrections assume $T_{e}$=4 eV, a typical value simulated in HPEM. Error bars correspond to 2 eV $<T_{e}<$6 eV. HPEM simulations required larger than expected applied voltages of 140 V in order to achieve reasonable densities. Very close agreement between experiment and simulation is observed until around 1 Torr. Above 1 Torr, grounded probe collision frequency remains lower than simulated values. A slightly logarithmic trend in $\nu_{en}$, as opposed to linear, is expected with increasing pressure since higher collisionality will shift the electron energy distribution function towards lower energies. This nonlinear trend is observed in figure 6(b) but is not clearly exhibited in figure 6(a). An explanation for this discrepancy is still an unresolved issue. The measured $\nu_{en}$ at higher pressures correspond to $T_{e}\approx$ 2.5 eV, while HPEM simulated $T_{e}\approx$ 3.7 eV at the location of the probe.

Additional measurements are taken over a range of gas compositions in Ar-He mixture plasmas and compared to HPEM runs at $p$=0.75 Torr and $P$=20 W. Sheath corrections assume $T_{e}$=4 eV. Error associated with uncertainty in $T_{e}$ is similar in magnitude to those in figure 6, but error bars were not included for illustrative purposes. Close agreement between measurement and simulation can be observed in figure 7. A notable increase in collision frequency from 90\% $\to$ 100\% He was observed in both experiment and simulation. This phenomenon is a result of a moderate increase in $T_{e}$ when transitioning to pure He. All mixing simulations were run at a slightly larger than expected applied voltage of 45 V (corresponding to $P$=32 W) in order to achieve reasonable densities, except for the 100\% He case which again required 140 V. Increasing the applied voltage typically decreased $T_{e}$ in the simulation. If the 100\% He case was capable of simulation at the self-consistent applied voltage of 45 V, one would expect simulation to yield a slightly more exaggerated increase in the 90\%-100\% He transition.

\begin{figure}[h]
\begin{center}
\includegraphics[scale=0.55]{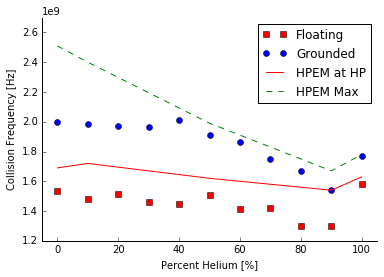}
\end{center}
\caption{Gas composition sweep of an argon-helium mixture with $p$=0.75 Torr and $P$=20 W. The sheath correction assumes $T_{e}$=4 eV.}
\end{figure} 

Reaction rates may be obtained from hairpin measurements of collision frequency using the ideal gas law, shown in equation (\ref{eq: reaction rate}). 
\begin{equation} \label{eq: reaction rate}
K(T_{e}) = \frac{T_{g}k_{B}\nu_{en}}{p}
\end{equation}
If this is used in conjunction with a tunable diode laser absorption measurement of neutral gas temperature ($T_{g}$), $T_{e}$ may be self-consistently determined with the sheath correction at conditions that are difficult to obtain with standard techniques. This can easily be done by comparing the measured reaction rate to expected reaction rates using BOLSIG+[15]. The method developed here may even be sensitive enough to infer $T_{e}$ in non-Ramsauer gases, albeit with less accuracy. Measurements are also amenable to time resolution using the boxcar method[13].

\section{Conclusions}

Collision frequency measurements are obtained using the hairpin resonance probe for Ar, He, and Ar-He mixture plasmas. Measurements match closely with simulation results, and have a maximum of approximately 20\% difference. Primary sources of error for this method stem from assuming electron temperatures and one Debye length sheaths. The presented technique presents an useful route for obtaining important information from plasmas at conditions that are becoming increasingly desirable for a number of industrial applications. This seems particularly useful for accurately determining electron temperature in moderate pressure plasmas considering the difficulty associated with using Langmuir probes at these pressures, and a notable lack of other available diagnostic techniques. Extending the technique developed here to determine $T_{e}$ is the subject of future work.

\section{Acknowledgements}
This work is supported through a generous gift by Applied Materials Inc. The author would like to thank Yiting Zhang for her invaluable help with HPEM simulations. 

\section{References\label{except}}

\clearpage

\end{document}